# Structural transition, metallization and superconductivity in quasi 2D layered $PdS_2$ under compression

Wen Lei,[1] Wei Wang,[1] Xing Ming,[1,*] Shengli Zhang,[2,†] Gang Tang,[3] Xiaojun Zheng,[1] Huan Li,[1] and Carmine Autieri[4]

1. College of Science, Guilin University of Technology, Guilin 541004, China.

2. MIIT Key Laboratory of Advanced Display Materials and Devices, School of Materials Science and Engineering, Nanjing University of Science and Technology, Nanjing 210094, China.

3. Theoretical Materials Physics, Q-MAT, CESAM, University of Liège, B-4000 Liège, Belgium.

4. International Research Centre MagTop, Institute of Physics, Polish Academy of Sciences, Aleja Lotników 32/46, PL-02668 Warsaw, Poland.

**ABSTRACT**

Based on first-principles simulations and calculations, we explore the evolutions of crystal structure, electronic structure and transport properties of quasi 2D layered $PdS_2$ under compression by uniaxial stress and hydrostatic pressure. An interesting ferroelastic phase transition with lattice reorientation is revealed under uniaxial compressive stress, which originates from the bond reconstructions of the unusual $PdS_4$ square-planar coordination. By contrast, the layered structure transforms to a 3D cubic pyrite-type structure under hydrostatic pressure. In contrary to the experimental proposed coexistence of layered $PdS_2$-type structure with cubic pyrite-type structure at intermediate pressure range, we predict that the compression-induced intermediate phase showing the same structure symmetry with the ambient phase, except of sharply shrinking interlayer-distances. The coordination of the Pd ions not only plays crucial roles in the structural transition, but also lead to electronic structure and transport properties variations, which changes from square-planar to distorted octahedron in the intermediate phase, resulting in bandwidth broaden and orbital-selective metallization. In addition, the superconductivity in the cubic pyrite-type structure comes from the strong electron-phonon coupling in presence of topological nodal-line states. The strong interplay between structural transition, metallization and superconductivity in $PdS_2$ provide a good platform to study the fundamental physics of the interactions between crystal structure and transport behavior, and the competition or cooperation between diverse phases.

---

[*] Email: mingxing@glut.edu.cn (Xing Ming)
[†] Email: zhangslvip@njust.edu.cn (Shengli Zhang)

## I. INTRODUCTION

Layered transition metal dichalcogenides (TMDs) $MX_2$ (M and X denote transition metal cation and chalcogen anion, respectively) have been studied intensively for several decades due to their widely technological applications and fundamental importance. Most of the bulk of $MX_2$ materials are quasi two-dimensional (2D) and consist of 2D layers bonded through van der Waals (vdW) interlayer interactions.[1] Recently, the TMDs materials $PdS_2$ and $PdSe_2$ with a novel layered structure have attracted extensive interests.[2-22] As shown in **Fig. 1 (a)**, at ambient conditions, bulk $PdX_2$ (X = S or Se) crystallizes in an orthorhombic space group (SG) $P$bca, named as $PdS_2$-type structure, consisting of unusual dumbbell shaped $(X_2)^{2-}$ anions and square-planar $(PdX_4)^{2-}$ structural units.[23] The layered structure can be viewed as orthorhombic distortion of the cubic pyrite structure (space group $Pa\bar{3}$) [**Fig. 1 (b)**], and the unusual $PdX_4$ square-planar coordination is originated from the distorted $PdX_6$ octahedron by elongation of the $c$ axis.[6] Furthermore, differed from common 2D monolayer materials, monolayer $PdX_2$ possesses novel pentagonal structures, which have been extensively studied due to their excellent performance, such as good air stability, high anisotropic carrier mobility, enhanced thermoelectric properties, and so on.[2,10-22]

In addition to the overwhelming research on the monolayer and few-layered phase, pressure or stress (strain) becomes an important tool to modulate the crystal structure, electronic structure and transport properties of $PdX_2$. Theoretical calculations show that strains play an important role in tuning the band gap, mechanical response, electronic structure and optical properties of monolayer $PdSe_2$, and half-metallic ferromagnetism can be realized in the monolayer $PdSe_2$ by hole-doping with applied uniaxial stress.[12,16,22] A new 2D high-pressure phase of $PdSe_2$ with high-mobility transport anisotropy has been theoretically predicated for photovoltaic applications.[24] Especially, ferroelastic phase transition accompanied with semiconductor-metal-semiconductor transitions has been predicted recently in pristine bulk $PdSe_2$.[2] Distinguished from common 2D ferroelastic materials, the ferroelastic phase transition in bulk $PdSe_2$ results from the reconstruction of Pd-Se bonds. The unusual stress-engineered properties endow the orthorhombic $PdSe_2$ a prospective material for potential applications in shape memory and other nanoscale devices. According to the pressure homology principle, similar ferroelastic structural transition can also be expected in $PdS_2$ under uniaxial stress.[6,25] Interestingly, structural transitions from the orthorhombic layered structure to cubic pyrite-type structure are experimentally realized under hydrostatic pressure in the pristine bulk PdSe$_2$ and PdS$_2$ (**Fig. 1**).[4-6,26] Raman spectra suggest that the PdS$_2$-type and pyrite-type phases coexist in the intermediate pressure range.[4-6]

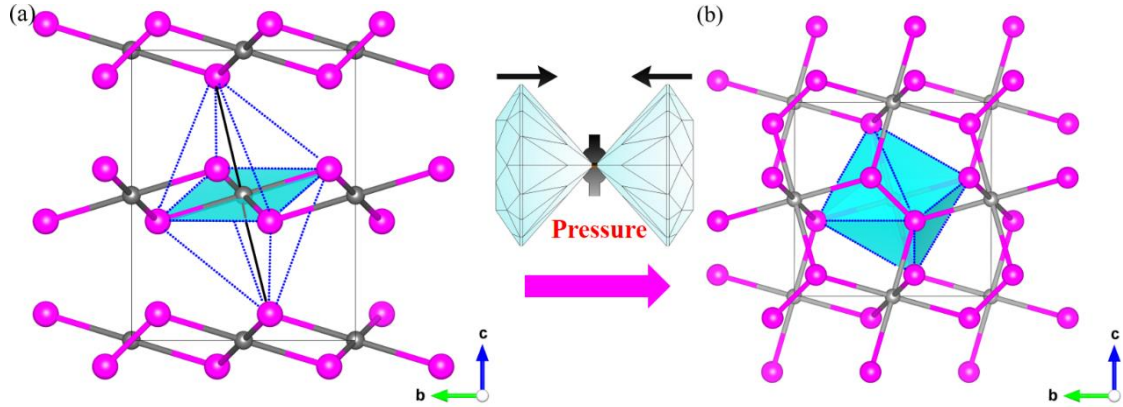

**Fig. 1** Illustration of the pressure-induced structural transition in PdX$_2$ from (a) the layered orthorhombic PdS$_2$-type structure to (b) 3D cubic pyrite-type structure. Big pink balls represent X atoms and small gray balls represent Pd atoms, respectively. The structural detail is highlighted by the X$_2$ anion dumbbells (red solid lines), whereas the PdX$_4$ square-planar and hidden distorted PdX$_6$ octahedra are plotted in cyan in (a) and (b), respectively.

More interestingly, the transport behaviors of bulk PdSe$_2$ and PdS$_2$ are similar to each other under high pressure. Electrical resistivity experiments demonstrate that application of hydrostatic pressure can tune the transport behavior of PdSe$_2$ (PdS$_2$) from semiconducting to metallic at pressures above 3 (7) GPa.[4,6] Furthermore, the pressure-induced metallization in PdX$_2$ occurs earlier than the structural phase transition to the cubic pyrite phase. Therefore, the metallization of the intermediate phase in PdX$_2$ is unassociated with the cubic pyrite-type structure, which has been proposed to originate from the filling-changes of the Pd-*d* orbitals, and named as orbital selective metallization.[6] In addition, superconducting states emerge in the high-pressure cubic pyrite-phase PdX$_2$ at low temperature, and the critical temperatures Tc show dome-shaped dependence on pressure.[4,6] Previous experiments propose that the superconducting transition temperature in the pyrite-phase PdSe$_2$ correlates with the weakening Se-Se bonds in the (Se$_2$)$^{2-}$ dumbbells.[4] However, although possessing similar Se-Se bond instability under pressure, the isoelectronic and isostructural pyrite-phase NiSe$_2$ remains a normal metal up to 50 GPa.[6] Moreover, high-pressure Raman spectra also do not show any anomalies due to the softening of the S-S bonding in pyrite-phase PdS$_2$. Therefore, the structural instability of the (X$_2$)$^{2-}$ dumbbells is not the necessary condition for the emergence of superconductivity in the high-pressure pyrite-type compounds. The Dirac and nodal-line fermions around the Fermi level and strong electron-lattice coupling might be viewed as the origin triggering the emergence of superconductivity in the high-pressure pyrite phases $PdX_2$.[4,6]

Motivated by the pressure or stress modulated mechanical properties, electronic structure and transport properties in bulk $PdSe_2$, we turn our attention to its isostructural and isoelectronic compound $PdS_2$ and explore the structure response and electronic properties under deformation by applying uniaxial compressive stress and hydrostatic pressure. Furthermore, we want to get further insight into the metallization in the intermediate phase and superconductivity in the high-pressure cubic pyrite-phase. We not only discover an interesting ferroelastic lattice reorientation under uniaxial stress, but also successfully reproduced the experiments observed structural transition from $PdS_2$-type to pyrite-type structure under hydrostatic pressure. Especially, the electronic structure and transport properties are tuned by the compression, which are closely correlated with the structure deformation and the coordination environments. The compression-induced metallization is ascribed to the bandwidth broaden and orbital population variation, whereas the superconductivity in the high-pressure pyrite phase originates from the strong electron-phonon coupling concomitant with topological nodal-line states.

The rest of this paper is organized as follows: **Sec. II** briefly describes the computational methodology. In **Sec. III**, we reveal ferroelastic transition under uniaxial compressive stress and study on structural transition from $PdS_2$-type to pyrite-type structure under hydrostatic pressure, then present analyses of the metallization of $PdS_2$-type phase under compression and superconductivity in the cubic pyrite phase. Finally, **Sec. IV** offers conclusions.

## II. COMPUTATIONAL AND ANALYTICAL METHODS

Geometry optimization and electronic structure calculations are carried out using first-principles methods based on density functional theory (DFT) as implemented in CASTEP[27] and VASP[28] code. The generalized gradient approximation in the Perdew-Burke-Ernzerhof[29] form of the exchange and correlation (XC) functional is used. In order to consider the layered structure and vdW interactions in the orthorhombic $PdS_2$, we use semiempirical dispersion interaction correction methods (DFT-D)[30] of Grimme scheme[31]. Recognizing that traditional XC functionals often underestimate the bandgap value, Heyd-Scuseria-Ernzerhof (HSE06)[32,33] hybrid density functional is used to calculate the electronic structure within VASP code.[28] The kinetic-energy cutoff is set up to 500 eV and the $k$-point spacing of the reciprocal space is fixed to 0.03 Å$^{-1}$. The integral calculation for the Brillouin zone uses the Monkhorst-Pack scheme to select the *k*-point grid. The phonon spectra of the bulk PdS$_2$ calculated by using a 2×2×2 supercell with the PHONOPY code on the basis of density functional perturbation theory.[34] Electron-phonon coupling were calculated by QUANTUM ESPRESSO[35] package with 8×8×8 *q* points and denser 16×16×16 *k* mesh.

In order to simulate the experimentally mechanical loading process, we specify an external stress tensor along the layer stacking direction (*c* axis) to probe the response of the crystal structure under uniaxial compression. The internal stress tensor is iterated until it becomes equal to the applied external stress, and the lattice constants gradually relax along with the iteration. Here, the produced deformation is defined as strain, $\varepsilon = (a - a_0)/a_0$, where $a$ and $a_0$ are the lattice constants along one axis direction for the deformed and undeformed cell, respectively. It is noteworthy that the simulation method in presented work is different from the conventional technique used in 2D materials, where the strain is independent variable, and is simulated by predesignated lattice constants rather than the stress.[8,12,16,22] We have calculated the evolutions of strain, lattice parameters, and energy along with applied uniaxial compressive stress ranging from 0 to 1 GPa with a step size of 0.05 GPa. To simulate the hydrostatic pressure condition, the external pressure is fixed and the cell volume at that pressure is relaxed by carrying out geometry optimization within CASTEP code.[27] All the geometric structures (including lattice constants and atomic internal coordinates) are fully relaxed. The convergence thresholds for maximum stress, and maximum displacement between optimization cycles are 0.02 GPa and $5 \times 10^{-4}$ Å.

Attributing to the layered structure and interlayer vdW interactions in PdS$_2$, the semiempirical dispersion interaction correction methods (DFT-D) (Grimme scheme)[30,31] has been used to calculate the electronic structure, optical and piezoelectric properties in monolayer and bulk phase PdS$_2$ (SG *P*bca).[8,13] **Table S1** and **Table S2** present our theoretical calculated lattice constants, bond lengths and the interlayer distances at ambient conditions, together with experimentally measured and theoretically calculated results.[36] The out-of-plane lattice constant *c* calculated without vdW correction is significantly larger than the experimental value, which indicates the significance of the vdW interactions in the titled material. The lattice constants and the electronic properties of bulk PdSe$_2$, isostructural to PdS$_2$, is also very sensitive to the choice of the XC functionals.[2,3,16] Compared with DFT-D method of Tkatchenko-Scheffler (TS) scheme[37], calculated results by Grimme scheme are closer to the experimentally measured values for PdS$_2$, which confirm the validity of our choice with the DFT-D (Grimme) method.

### III. RESULTS AND DISCUSSIONS

**A. Ferroelastic transition under uniaxial compressive stress**

Firstly, when the orthorhombic layered PdS$_2$ is subject to uniaxial compressive stress along the *c*-axis, PdS$_2$ transforms to an intermediate phase and finally recovers to the initial structure accompanied with the stacking direction of the puckered 2D PdS$_2$ layers rotating by 90° from *c*-axis to *b*-axis. PdS$_2$ maintains the layered structure with the same *P*bca SG after the phase transition. As shown in **Fig. 2(a)**, the uniaxial compressive stress induces out-of-plane longitudinal contraction and in-plane lateral expansion, resulting in out-of-plane compressive strain ($\varepsilon_c$, negative values) in the layer stacking direction and in-plane tensile strains ($\varepsilon_a$ and $\varepsilon_b$, positive values). At the same time, the out-of-plane lattice constant *c* obviously reduces under uniaxial compression, along with the intralayer lattice constants *a* and *b* slight increase [**Fig. 2(b)**]. The out-of-plane Pd-S distances (d1) gradually shrink, while the in-plane Pd-S bonds (d2 and d3) in the square-planar (PdS$_4$)$^{2-}$ structural units and the S-S bonds (d4) in the (S$_2$)$^{2-}$ anions are almost insensitive to the compression. The (PdS$_4$)$^{2-}$ square-planar structural units keep unchanged up to 0.55 GPa. However, when the compressive stress exceeds 0.55 GPa, the out-of-plane compressive strain $\varepsilon_c$ abruptly changes to -14.76%, and the in-plane tensile strains $\varepsilon_a$ and $\varepsilon_b$ obviously increase [**Fig. 2(a)**]. At the same time, the out-of-plane lattice constant *c* remarkably shrinks and the in-plane lattice constants abruptly expand [**Fig. 2(b)**], leading to the crystal structure transform to an intermediate phase. The intermediate phase still maintains the identical crystallographic symmetry (SG *P*bca) with the ambient-pressure phase, but the lattice constant *c* is significantly reduced. As shown in **Fig. 2(c)**, the in-plane Pd-S bonds (d2 and d3) in the square-planar (PdS$_4$)$^{2-}$ structural units are suddenly stretched in the intermediate phase, whereas the out-of-plane Pd-S distance (d1) is markedly compressed. Because of the obvious reducing interlayer-distances between the Pd and S atoms, the Pd-centered coordination environment changes from square-planar to elongated octahedron, but cubic crystal structure is not obtained. By contrast, the isostructural PdSe$_2$ can transform to a cubic pyrite-type structure (SG $Pa\bar{3}$) under uniaxial compressive stress.[2]

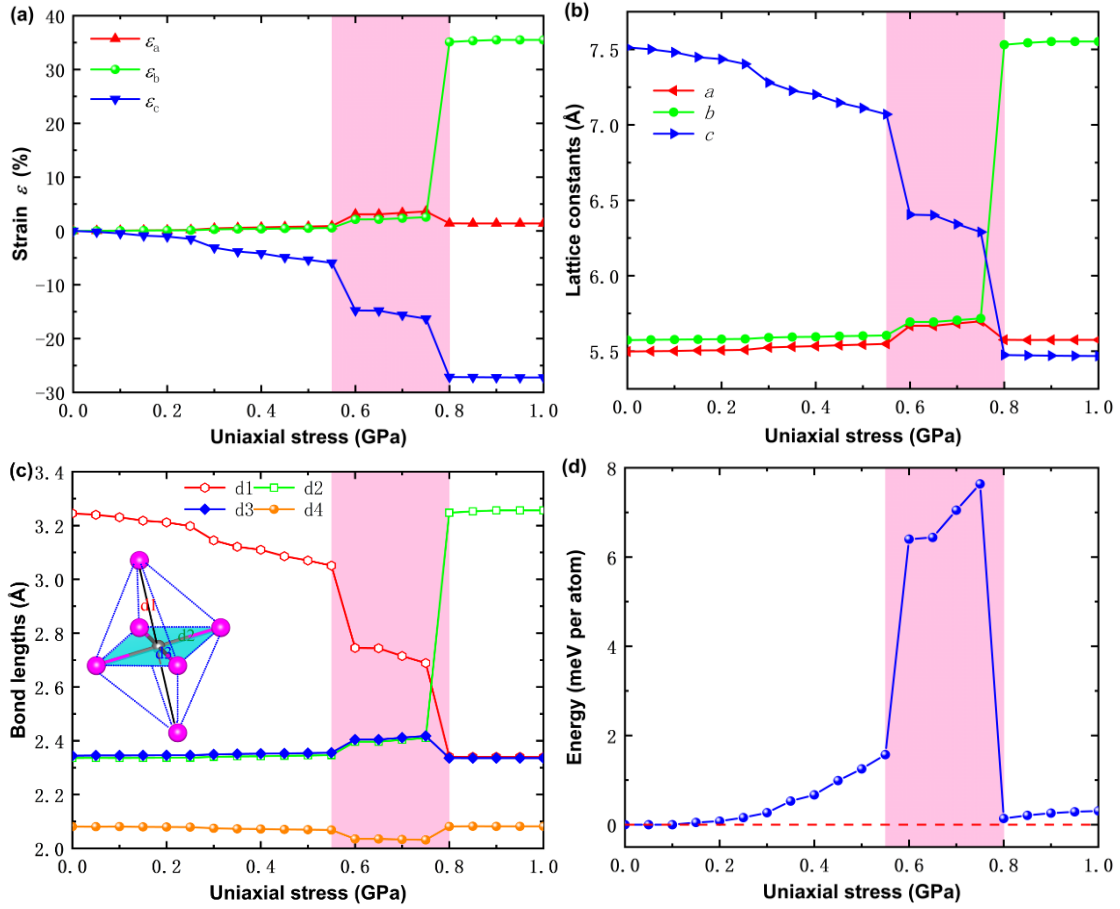

**Fig. 2** (a) Strain, (b) lattice constants, (c) bond lengths and (d) energy evolutions in response to the uniaxial compression along the *c*-axis of the orthorhombic phase PdS$_2$. The insets in (c) display the Pd-S inter-atomic distances in the distorted PdS$_6$ octahedron. d1 denotes the interlayer distance between Pd and S atoms, d2 and d3 are the Pd-S bond lengths in the square-planar (PdS$_4$)$^{2-}$ structural units, d4 is the S-S bond lengths in the (S$_2$)$^{2-}$ anions.

Further increasing the compressive stress to 0.8 GPa, the out-of-plane strain $\varepsilon_c$ drastically changes to -27.14%, while the in-plane strain $\varepsilon_b$ abruptly increases to 35.12%. At the same time, the in-plane Pd-S bonds (d2) suddenly increase and the out-of-plane Pd-S distance (d1) abruptly decreases. As a consequence, the bonding configurations achieve a simultaneous change with in-plane old-bond breaking and out-of-plane new-bond emerging, which causes the layer stacking direction to rotate from *c*-axis to *b*-axis. The ferroelastic lattice reorientation accompanied by bond reconstruction can be further inspected from the electron density differences as shown in **Fig. 3**. For the initial phase at ambient pressure [**Fig. 3(a)**], the electrons transfer from Pd atoms to four nearest neighbor S atoms in the layered (PdS$_4$)$^{2-}$ plane, the electron density differences show an obviously anisotropic feature, which is consistent with the bonding configurations of the square-planar coordination. Although the interlayer distances obviously reduce in the intermediate phase, the main bonding is still in the layered *ab* plane, but notable electrons transferring and bonding appears between the $(PdS_4)^{2-}$ layered plane [**Fig. 3(b)**], agreeing well with the elongated octahedral coordination of the intermediate phase. For the final phase at 0.8 GPa, the in-plane distributions of electron density difference obviously rotate 90° with respect to the unstrained case [**Fig. 3(a)**], implying the formation of new Pd-S covalent bonds in the *ac* plane [**Fig. 3(c)**]. The electron density differences confirm the simultaneous bond breaking and forming during the ferroelastic phase transition.

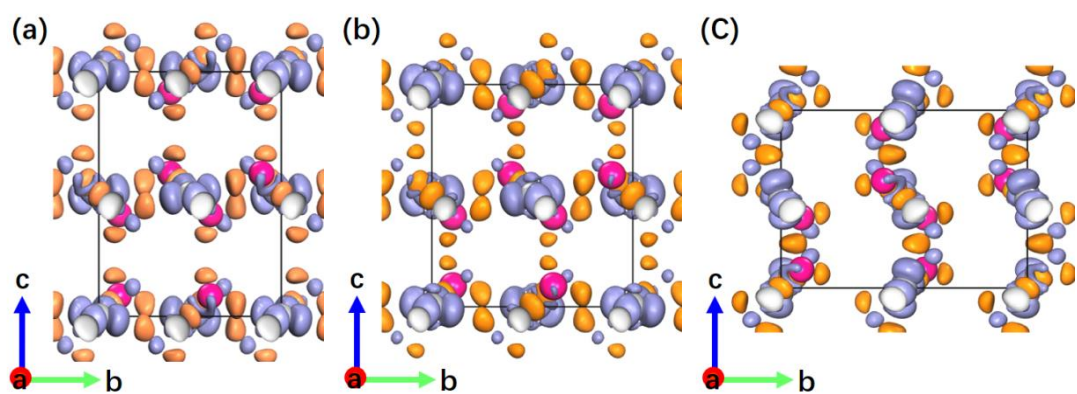

**Fig. 3** Electron density differences for the three stages of the ferroelastic phase transition: (a) initial phase at 0 GPa, (b) intermediate phase at 0.7 GPa, (c) final phase at 0.8 GPa. The isosurface value is taken as 0.1 e/Å$^3$ (light blue and orange colors indicate the regions of electron depletion and electron enrichment, respectively), and the pink/gray balls represent S/Pd atoms, respectively.

The ferroelastic phase transition can be further inspected from an energetic viewpoint. As shown in **Fig. 2(d)**, the energy variation shows a quadratic increase within the elastic region up to 0.55 GPa. Beyond the elastic region above 0.55 GPa, accompanied with the phase transition to the intermediate phase, the energy abruptly increases by about 4.83 meV/atom. In addition, the enthalpy linearly increases along with the increasing stress before the transition, and then suddenly increases with a barrier and the corresponding unit cell volume collapse by about 7% to the intermediate phase (**Fig. S1**).[36] The intermediate phase is metastable and spontaneously transforms into a stable configuration when the stress further increases to 0.8 GPa. Accompanied by the ferroelastic lattice reorientation, the energy recovers to the value of the original unstrained phase with an interchange of the *c* axis and *b* axis. The reversible ferroelastic strain (34.8%) (defined as $(c/b - 1) \times 100\%$) is close to the predicted values in the isostructural PdSe$_2$ (32.3%)[2] and other 2D monolayer material such as phosphorene (37.9%)[38] and BP$_5$ (41.41%)[39]. Furthermore, compared with previously reported 2D monolayer ferroelastic material with switching energy barriers ranging from 0.8 to 320 meV/atom,[38-44] the energy barrier in bulk PdS$_2$ is moderate (7.64 meV/atom) (**Fig. 4** and **Table S3**).[36] Generally, larger ferroelastic strain often coexists with higher energy barrier in 2D monolayer materials, indicates that it is hard to realize the ferroelastic switching in experiments.[41,43] A low energy barrier implies it is feasible to experimentally realize the ferroelastic switching and lattice reorientation in PdS$_2$.

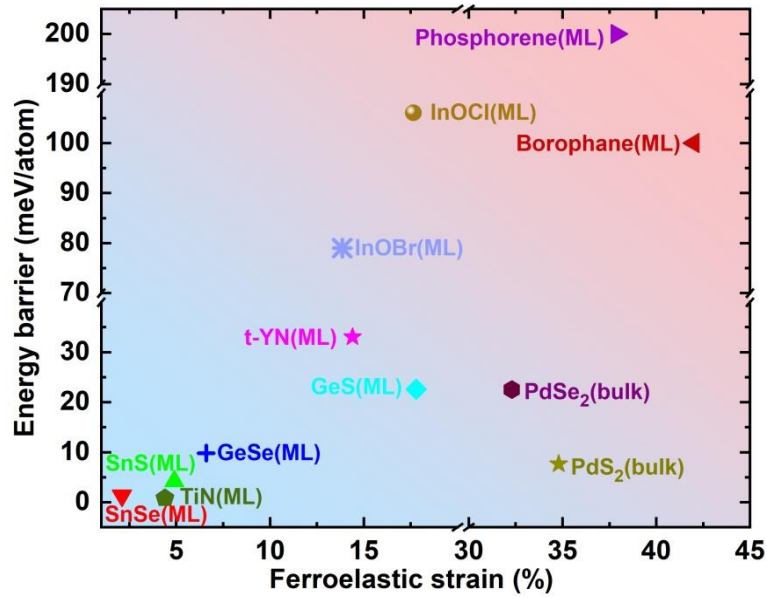

**Fig. 4** The activation energy barrier *vs*. ferroelastic strain in different materials (ML stands for 2D monolayer).

Due to potential applications in nonvolatile memory devices, two-dimensional (2D) ferroelastic/ferroelectric materials have been widely explored recently.[38-44] These 2D ferroelastic/ferroelectric materials undergo lattice orientation transitions under external stress or electric field.[45] However, many theoretically predicted 2D ferroelastic materials are not stable under environmental conditions, and large ferroelastic strain often coexists with high switching energy barrier, which hinders the practical application of ferroelastic materials in nonvolatile memory devices.[41,43] Furthermore, although plenty of 2D ferroelastic materials have been theoretically predicted, their monolayer polymorphs are yet to be fabricated.[46] Therefore, if the ferroelasticity can be realized in already-synthesized materials, especially in the quasi-2D materials with layered structure, their direct applications into nanoscale devices are more feasible.[2,46]

Intriguingly, attributed to the unusual crystal structure units of the $(PdX_4)^{2-}$ square-planar and the $(X_2)^{2-}$ anions, low transformation stress and moderate energy barrier coexist with an ultrahigh reversible ferroelastic strain in the isostructural $PdS_2$ and $PdSe_2$. These two layered TMDCs materials feature low transformation stresses and switching barriers along with strong ferroelastic signals, renders the desirable possibility of fast ferroelastic switching upon imposing external stress, beneficial for practical applications in flexible electronics and shape memory devices.[38,39]

**B. Structural transition to the cubic pyrite-type structure under hydrostatic pressure**

As shown in **Fig. 1**, recently experiments reveal structural transition from the layered orthorhombic structure to cubic pyrite-type structure under high-pressure in $PdS_2$.[6] The pressure-induced structural transition in $PdS_2$ has been reproduced by our theoretical calculations. The lattice constants of the orthorhombic phase show highly anisotropic compressive behavior. As shown in **Fig. 5**, the lattice constant $c$ rapidly decreases along with the compression before 1 GPa, whereas the lattice constants $a$ and $b$ increase slowly. The shrinkage of the lattice constant $c$ originates from the collapse of the interlayer spacing. The compressibility is consistent with the quasi-2D layered structure stacking along the crystallographic $c$ axis, deriving from the weak interlayer vdW interactions. Upon the hydrostatic pressure further increasing, the crystal structure transforms to an intermediate phase, which maintains the layered structure with the same SG of $P$bca and structural symmetry. However, the coordination environments of the Pd atoms change from square-planar to distorted octahedron due to the reducing interlayer Pd-S distance (d1). Along with further compression, the out-of-plane lattice constant $c$ gradually shrinks and the in-plane lattice constants further expand. Once the pressure reaches to 4 GPa, the lattice constants gradually become to the same with each other, and the interlayer Pd-S distances (d1) are equal to the Pd-S bond lengths (d2 and d3) in the $(PdS_4)^{2-}$ square-planar structure. As a consequence, the layered orthorhombic $PdS_2$-type structure transforms to the cubic pyrite structure.

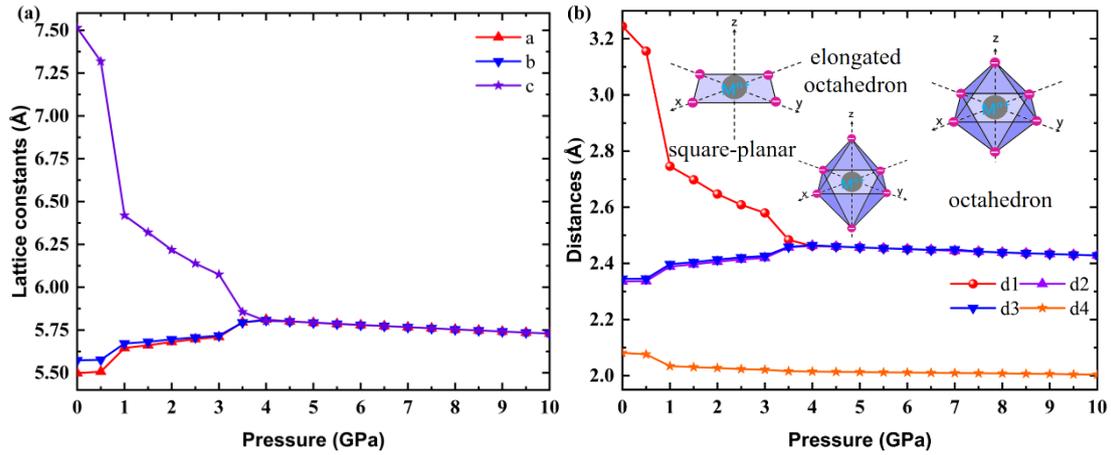

**Fig. 5** Variation of the lattice parameters of $PdS_2$ in response to the hydrostatic pressure: (a) lattice constants and (b) inter-atomic distances (d1, d2 and d3) in the distorted $PdS_6$ octahedra and S-S bond length (d4) in the $(S_2)^{2-}$ dumbells. The insets in (b) display the transformation of the coordination polyhedra from square-planar to octahedron.

High-pressure experiments show three remarkable stages of the structural transition in $PdX_2$. Especially, the $PdS_2$-type structure is proposed to coexist with the cubic pyrite-type structure for the intermediate phase.[4-6] Our theoretical simulation confirms the existence of an intermediate phase in $PdS_2$ at high pressure. However, the theoretical predicted results suggest that the intermediate phase is not a coexisting phase but still an orthorhombic structure. Experimentally, metastable high-pressure Pd deficient phases with the SG *P*bca have been synthesized via high-pressure/high-temperature method by Gunter Heymann *et al*, where the lattice constant *c* significant decreases compared to the ambient-pressure phase.[47] As shown in **Fig. S2**,[36] the calculated phonon spectrum does not show any imaginary frequencies and soft acoustic branches, indicating that the dynamical stability of both the layered orthorhombic $PdS_2$-type structure at ambient condition and the cubic pyrite structure under high pressure. Furthermore, the intermediate phase is also confirmed to be dynamical stable by the phonon dispersion curve without imaginary frequencies (**Fig. S3**).[36] Therefore the pressure-induced structural transition is not originated from the dynamical instability.

The nature of the structural transition of $PdSe_2$ from the $PdS_2$-type to cubic pyrite-type has been proposed to be displacive.[5] According to the pressure homology principle, the origin of the pressure-induced structural transition in $PdS_2$ can also be expected to be the same with the isostructural $PdSe_2$. As shown in **Fig. 1**, the $PdS_2$-type crystal structure consists of 2D layers with square-planar coordinated Pd ions. The out-of-plane ions gradually approach the coordinated-central

Pd ions under compression, and forming distorted $PdS_6$ octahedra elongated along the *c* axis.[4-6] The elongation of the octahedron is belonged to Jahn-Teller (JT) deformation, which is derived from the degenerate state of the *d*-electron configuration. The distortion degree can be estimated by the distortion degree parameter $\Delta d = 1/6\sum_{i-1}^{6}[(R_i - \bar{R})/\bar{R}]^2$, where $R_i$ and $\bar{R}$ denote individual interatomic Pd-S distance and average distance, respectively.[48] As shown in **Fig. S4**,[36] the distortive degree ($\Delta_{pd}$) as a function of pressure decreases rapidly before 1 GPa. The interlayer and intralayer Pd-S distances are about 3.2 and 2.3 Å at ambient condition. The vdW radii of Pd and S ions are 2.15 and 1.8 Å, whereas the covalent radii of Pd and S ions are 1.39 and 1.05 Å, respectively.[49,50] Apparently, the interlayer Pd-S distances are on the verge of vdW bond (3.1~5.0 nm). Further compression leads to the interlayer Pd-S distance below the threshold value of vdW interactions. The collapse of the interlayer spacing indicates the interlayer vdW interactions annihilation and covalent bond appearing. Consequently, the quasi-2D $PdS_2$ transforms to a new intermediate phase with distorted $PdS_6$ octahedron. The distortive degree of the $PdS_6$ octahedron is gradually suppressed up to 4 GPa, and the crystal structure changes to the cubic pyrite-type. The compression eliminates the JT structural distortion in $PdS_2$ and renders the pyrite-type structure stable under high-pressure.[4,6]

## C. The metallization of orthorhombic phase under compression and superconductivity in the cubic pyrite phase

Under ambient conditions, the orthorhombic $PdS_2$ exhibits semimetallic behavior at the DFT level (**Fig. 6**), which is consistent with previous calculations.[7] But resistivity measurements demonstrate that $PdS_2$ is nonmetallic with energy gap of 0.7-0.8 eV.[51] Therefore we employ HSE06 screened Coulomb hybrid density functional to calculate the band structure and an indirect band gap of 0.818 eV opens up [**Fig. 6(a)**]. Similar to the electronic structure of the isostructural $PdSe_2$,[2] spin-orbit coupling interactions almost play no impact on the electronic structure of $PdS_2$ (**Fig. S5**).[36] As shown in **Fig. 6(b)**, under uniaxial compression, the band gap gradually decreases before transforming to the metallic intermediate phase (detailed band structures are shown in **Fig. S6-S7**).[36] When the ferroelastic phase transition is realized under uniaxial compression, the orthorhombic $PdS_2$ recovers to the semiconducting property (**Fig. S7-S8**).[36] The evolutions of the electronic structure are closely related to the interlayer spacing (**Fig. S9**).[36] Compared with other chemical bonds, the vdW force is a weak interaction, and the electronic structure is very susceptible to the external environment and interlayer coupling.[52-56] As shown in **Fig. 6(a)**, the valence band maximum (VBM) and conduction band minimum (CBM) mainly originate from the hybridization of S 3$p$ and Pd 4$d$ states in orthorhombic PdS$_2$, which is similar to the case of orthorhombic PdSe$_2$[2,4] and other TMDCs[7,52]. Because of different orbital contributions, the influences of the interlayer coupling on VBM and CBM are very different.[57] Under uniaxial compressive stresses along the layer stacking direction, the interlayer spacing reduce and the interlayer coupling enhance, resulting in obvious impact on the state near the band edge and inducing metallization.[58] Once the layer stacking direction has switched from $c$-axis to $b$-axis after the ferroelastic phase transition, the interlayer distances become larger and larger with increasing stress (**Fig. S9**).[36] As a result, the larger the interlayer distance the weaker the interlayer interaction and the weaker the interlayer coupling effect, then metal to semiconductor transition occurs. Therefore, the band gap can be modulated by the interlayer spacing through uniaxial stress or other mechanical deformations. The geometric, mechanic and electronic properties are sensitive to the stress or strain, implying the layered orthorhombic PdS$_2$ and PdSe$_2$ can be applied to flexible electronics, nanomechanics and sensor devices.[2,59]

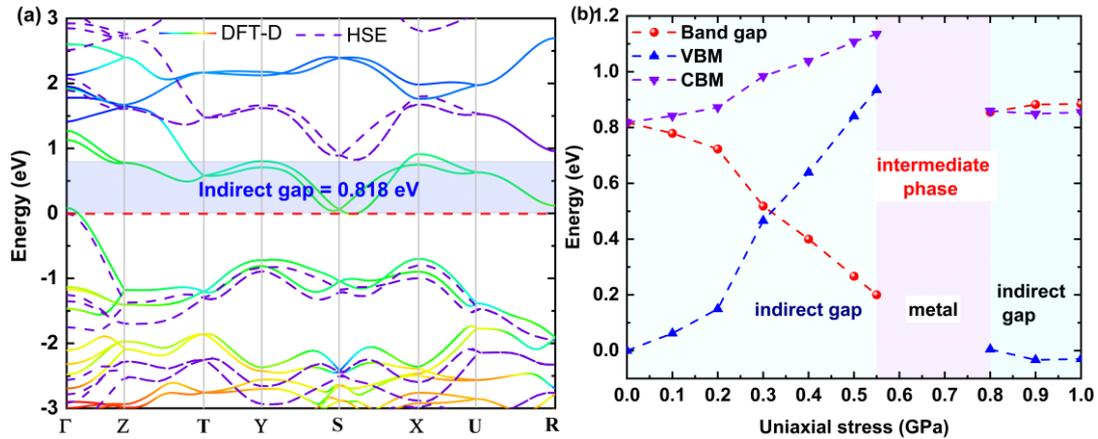

**Fig. 6** (a) Band structure of orthorhombic PdS$_2$ under ambient condition calculated with DFT-D (Grimme) and HSE06 functionals. The color denotes the contributions from Pd-4$d$ (red) and S-3$p$ (blue) states, and strong hybridizations between them. (b) The evolution of band edge and band gap under uniaxial stress calculated within HSE06. The Fermi level and VBM under ambient condition are set to 0 and as the reference point.

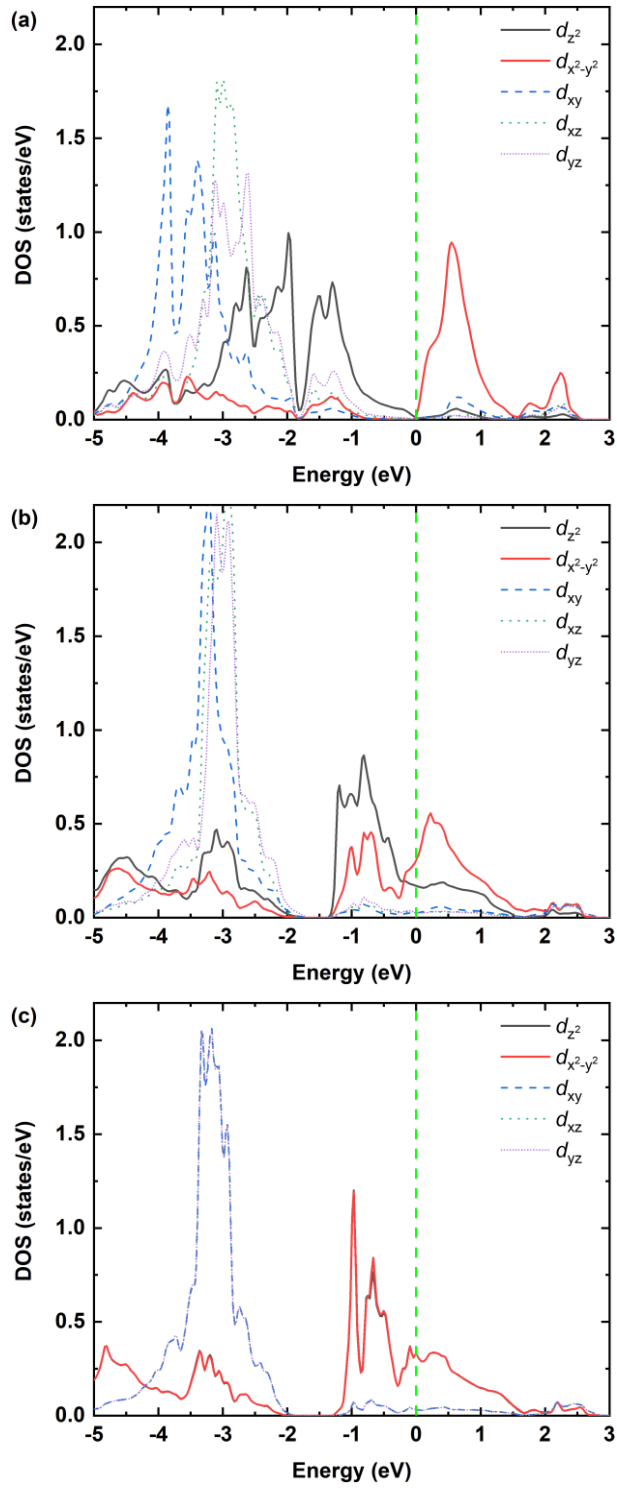

**Fig. 7** Projected density of states (PDOS) evolutions of the Pd 4$d$ state in PdS$_2$ for different stages of the phase transitions under compression: (a) initial phase at 0 GPa, (b) intermediate phase at 2 GPa and (c) high-pressure cubic pyrite-type phase at 4 GPa. The Fermi level is set to zero.

In addition, electrical resistivity experiments demonstrate that application of hydrostatic pressure also can tune the transport behavior of PdS$_2$ from semiconducting to metallic at pressures above 7 GPa.[6] Similarly, the room-temperature resistivity show semiconductor-to-metal transition at 3 GPa in the isostructural PdSe$_2$.[4] Furthermore, the pressure-induced metallization in PdX$_2$ occurs earlier than the structural phase transition to the cubic pyrite phase. Therefore, the metallization of the intermediate phase in PdX$_2$ is not related to the cubic pyrite-type structure, which has been proposed to originate from the filling-changes of the Pd-$d$ orbitals, and named as orbital selective metallization.[6] As we known, the $d$ orbitals are split into triply degenerate $t_{2g}$ states and doubly degenerate $e_g$ states in the octahedral crystal-field. The JT effect results in a further splitting of the $e_g$ levels in the distorted octahedral coordination. Especially, the energy of the $d_{z^2}$ orbital will drop down relative to the $d_{x^2-y^2}$ orbital when the octahedron is elongated along the $z$ axis. If the two coordinating anions go further away from the coordinated-central cations, the coordination environments will change from distorted octahedron to square-planar and the $d_{z^2}$ orbital drops to further lower energy range. The PdS$_2$-type structure consists of 2D layers of square-planar coordinated Pd ions at ambient condition (**Fig. 1**). Therefore the eight electrons of Pd$^{2+}$ ions fill all the $d$ orbitals except of the $d_{x^2-y^2}$ orbital, accounts for the diamagnetic semiconducting properties of PdX$_2$ at ambient conditions.

Detailed changes of the electronic structure can be found in the projected density of states (PDOS) joint with band structure. As shown in **Fig. 7(a)** and **Fig. 8(a)**, in good agreement with the crystal-field theory, for the uncompressed PdS$_2$ at ambient conditions, the Pd $d_{x^2-y^2}$ orbital mainly contributes to the bottom of the conduction bands, whereas the doubly occupied $d_{z^2}$ orbital strongly hybridizes with other $d$ orbitals locating at lower energy range. The interlayer Pd-S distances decrease sharply upon compression, no matter under uniaxial stress or hydrostatic pressure. As shown in **Fig. S10**,[36] although the compression leads to bandwidth strongly broadening, PdS$_2$ remains semiconducting until the structure changing to the intermediate phase. Once the PdS$_2$-type structure transforms to the intermediate phase, the valence bands overlap with the conduction bands, leading to the semiconductor to metal transition. Essentially, the population-changes of the Pd-$d$ orbitals play important role in the metallization of PdS$_2$. Along with the structure transforming to the intermediate phase, the coordination field changes from square-planar to distorted octahedron. As shown in **Fig. 7(b)** and **Fig. 8(b)**, the crystal-field splitting between $d_{x^2-y^2}$ and $d_{z^2}$ orbitals reduces, where the empty $d_{x^2-y^2}$ orbitals mix with the $d_{z^2}$ orbitals. In addition to the broadening of the bands around Fermi level, the partial-filling of the $d_{x^2-y^2}$ orbitals obviously contribute to the metallization. Present theoretical calculation results provide direct evidences of orbital selective metallization in the intermediate phase.[6] Accompanied with increasing pressure and the structrual transition from orthorhombic intermediate phase to cubic pyrite phase, the distortion of the PdS$_6$ octahedron is almost eliminated except for slight deviations of the bond angles from 90°. The $d_{x^2-y^2}$ orbitals completely coincide with the $d_{z^2}$ orbitals, and the splitting between them disappears [**Fig. 7(c)**]. A clear octahedral crystal-field splitting appears between the triply degenerate $t_{2g}$ states ($d_{xy}$, $d_{xz}$ and $d_{yz}$) and doubly degenerate $e_g$ states ($d_{x^2-y^2}$ and $d_{z^2}$). The half-filled $e_g$ states results in the metallic transport properties in the cubic pyrite-type PdS$_2$ under high pressure.

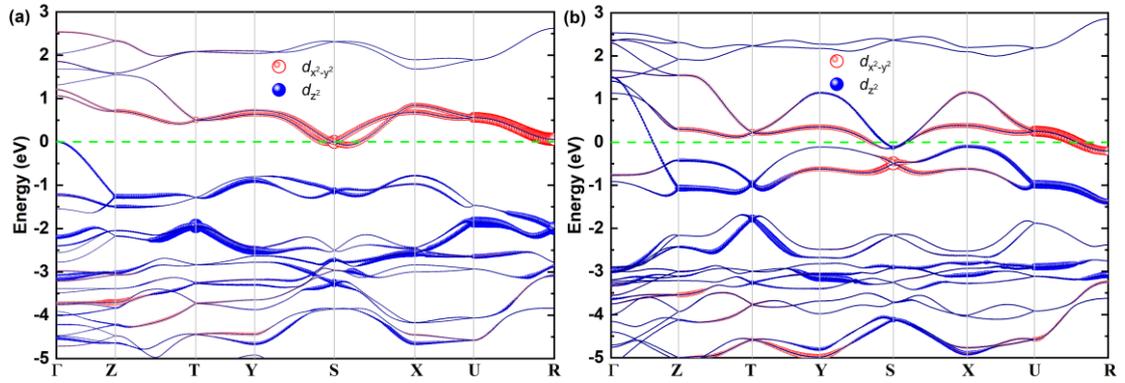

**Fig. 8** Electronic band structures of orthorhombic PdS$_2$ calculated with DFT-D (Grimme) for the: (a) ambient pressure phase, and (b) intermediate phase at 2 GPa. The Fermi level is set to zero. The contributions of the $d_{z^2}$ (blue solid balls) and $d_{x^2-y^2}$ (red hollow balls) orbitals are proportional to the size of the colored balls, as indicated in the legend.

More interestingly, the transport properties of PdS$_2$ are qualitatively similar to those of PdSe$_2$ in the pyrite-type phase under high pressure. Superconductivity emerges at low temperature, showing dome-shaped superconducting diagram.[4,6] Previous experiments propose that the superconducting transition temperature in the pyrite-phase PdSe$_2$ is associated with weakening of Se-Se bonds in the (Se$_2$)$^{2-}$ dumbbells.[4] However, although possessing similar Se-Se bond instability under pressure, the isoelectronic and isostructural pyrite-phase NiSe$_2$ remains a normal metal up to 50 GPa.[6] Furthermore, in contrast to the anomalies of the Se-Se bonding strengthen in the pyrite-phase PdSe$_2$ and NiSe$_2$, our theoretical calculations reveal that the S-S bond lengths of the (S$_2$)$^{2-}$ dumbbells in pyrite-phase PdS$_2$ are nearly independent on pressure [**Fig. 5(b)**]. High-pressure Raman spectra also not show any anomalies due to the softening of the S-S bonding in pyrite-phase PdS$_2$. Therefore, the structural instability of the (X$_2$)$^{2-}$ dumbbells is not the necessary condition for the emergence of superconductivity in the high-pressure pyrite-type compounds. The strong electron-lattice coupling in the high-pressure pyrite phases of $PdS_2$ might be viewed as the origin triggering the emergence of superconductivity.[6]

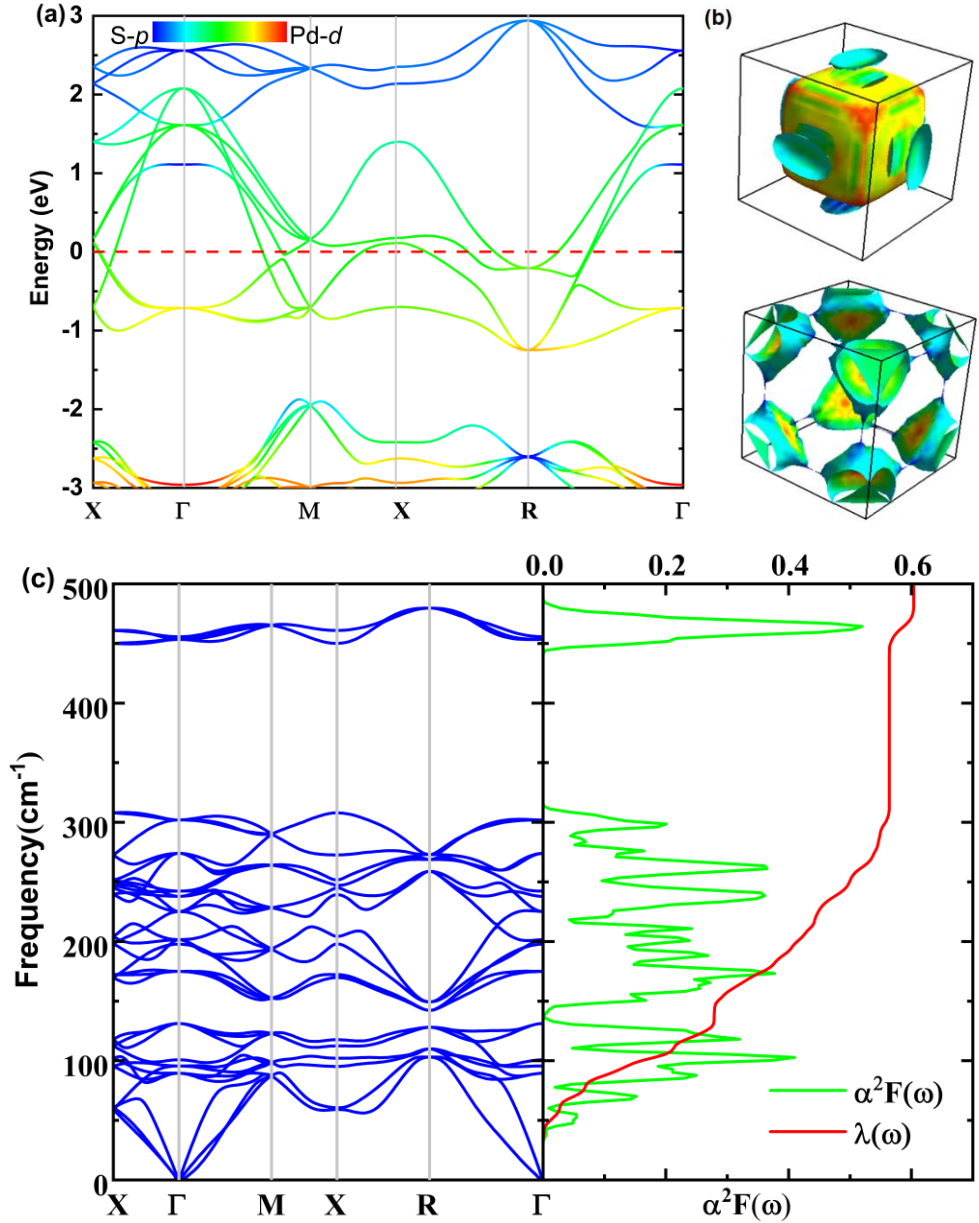

**Fig. 9** Electronic structure of the high-pressure pyrite-type $PdS_2$: (a) the band structure and (b) the corresponding Fermi surface. (c) Left: phonon dispersion curves for pyrite-type $PdS_2$ at 4 GPa. Right: Eliashberg electron-phonon coupling spectral function $\alpha^2F(\omega)$ and the electron-phonon coupling integral $\lambda(\omega)$.

In order to get further insights to the superconductivity in high-pressure pyrite-type $PdS_2$, we present the electronic structure in **Fig. 9**. The triply degenerate $t_{2g}$ states form localized narrow bands at lower energy region, which are separated from the half-filled $e_g$ states by a big gap of 0.7 eV. The bands cross over the Fermi level with strong hybridization between the S 3$p$ states and Pd 4$d$ states, exhibiting excellent metallic transport properties. Because of the inversion symmetry of the pyrite-type structure, all bands are doubly degenerate. At the same time, all of the special high-symmetry points in the Brillouin zone are time-reversal invariant. Therefore, there are nodal-line states along the Γ-X, Γ-M and X-M lines when SOC interactions are excluded. In addition, the band structure shows eightfold degeneracies (considering spin degeneracy) at the M point, and are fourfold degenerate along the M-X and X-R lines, which are protected by non-symmorphic symmetries.[60] The sixfold degeneracies at the R points are stem from the three-fold rotation symmetry.[61,62] In consistent with the nodal lines and the multifold degeneracies, the Fermi surface consists of a large cubelike pocket in the center of the Brillouin zone, and a network of small electron and hole pockets connected along the edges of the Brillouin zone [**Fig. 9(b)**]. The small curvature of the cubelike Fermi surface renders pyrite-type PdS$_2$ a strong candidate to produce quantum oscillations and large effective mass.[63,64] Recently, the superconducting instability has been proposed to interplay with the topological nontrivial states.[65,66] Electronic band structure calculations speculate that the superconducting states originating from the Dirac and nodal-line states in the vicinity of the Fermi level in the high-pressure pyrite-type PdSe$_2$.[4] Although the S-S bonding behavior of the (S$_2$)$^{2-}$ dumbbells in pyrite-phase PdS$_2$ differs from PdSe$_2$, similar superconductivity behaviors are observed, the superconducting phase PdS$_2$ and PdSe$_2$ feature similar dispersion characteristics of the band structure. Therefore, the topological nodal-line states are important for the emergent superconductivity in the pyrite-type PdX$_2$.

In addition, conventional superconductivity is closely associated with electron-phonon coupling. Further theoretical analyses of the electron-phonon coupling interactions are necessary to understand the superconducting behavior of the high pressure pyrite-type PdX$_2$. As shown in the left of **Fig. 9(c)** and **Fig. S11**,[36] the calculated phonon spectrum does not show any imaginary frequencies and soft acoustic branches, demonstrating that the dynamical stability of the pyrite-type PdX$_2$ under high pressure. Then, we inspect the superconducting properties of PdX$_2$ at high pressures by calculating the electron-phonon coupling constant $\lambda$ and superconductivity critical temperature Tc, using the Allen-Dynes-modified McMillan formula:[67]

$$T_c = \frac{\omega_{log}}{1.2} exp(-\frac{1.04(1+\lambda)}{\lambda-\mu^*-0.62\lambda\mu^*}), \quad (1)$$

where the Coulomb pseudopotential $\mu^*$ is assumed to 0.1, and the electron-phonon coupling constant $\lambda$ is defined as

$$\lambda = 2\int_0^\infty \frac{\alpha^2 F(\omega)}{\omega} d\omega, \quad (2)$$

and $\omega_{log}$ is the logarithmic average frequency:

$$\omega_{log} = exp(\frac{2}{\lambda}\int \frac{log\omega}{\omega}\alpha^2 F(\omega)d\omega), \quad (3)$$

where $\alpha^2 F(\omega)$ is the Eliashberg spectral function defined as

$$\alpha^2 F(\omega) = \frac{1}{2\pi N(E_f)}\Sigma_q \delta(\omega-\omega_q)\frac{\gamma_q}{\hbar\omega_q}, \quad (4)$$

where $N(E_f)$ is Fermi level and $\gamma_q$ is the phonon linewidth. Calculated results show that there are strong electron-phonon interactions in pyrite-type $PdX_2$, the electron-phonon coupling constant $\lambda$ of the pyrite-type $PdS_2$ ($PdSe_2$) at 4 (10) GPa is evaluated to 0.6 (0.9), and the corresponding superconducting transition temperature Tc is 4.9 (7.6) K. Present theoretically calculated superconducting temperatures are consistent well with previous experimental results.[4,6] The superconducting states emerge in pyrite $PdS_2$ ($PdSe_2$) at about 20.7 (7.2) GPa below 2 (2.4) K, and the critical temperature Tc increases up to maximum values of 8.0 (13.1) K at 37.4 (23) GPa. The superconducting temperatures are much higher than those of the extensively studied copper pyrites ($CuS_2$ and $CuSe_2$).[68] Although TMDCs compounds often crystallize in the so-called pyrite structure and exhibit various electronic states and magnetic properties, superconducting pyrite compounds are seldom reported so far. We have demonstrated that the pyrite-type $PdX_2$ show superconductivity with strong electron-phonon coupling, but this could also lead to instability towards more exotic superconductive orders. Therefore, the superconducting pyrite-type $PdX_2$ compounds are featured with topological nodal-line characteristics and nonsymmorphic symmetries, provide a promising platform to explore various intriguing electronic properties, such as topological superconductivity and topological nonsymmorphic crystalline superconductivity.[69,70]

## IV. CONCLUSIONS

In summary, the structure deformation and electronic property evolutions of the layered orthorhombic $PdS_2$ under uniaxial compression and hydrostatic pressure are systemically investigated by first-principles calculations. A ferroelastic lattice reorientation under uniaxial stress is uncovered, and the experiments observed structural transition from $PdS_2$-type to pyrite-type structure under hydrostatic pressure is also successfully reproduced. Especially, contrary to the experimental proposed intermediate phase of coexisting layered $PdS_2$-type structure with cubic pyrite-type structure, we predict that the compression-induced intermediate phase showing the same structural symmetry with the ambient phase due to sharply contracted interlayer-distances. Furthermore, the electronic structure and transport properties are closely related to the structural deformation and the coordination environments of the Pd ions. Along with the appearance of the intermediate phase, the coordination polyhedron changes from $PdS_4$ square-planar to distorted $PdS_6$ octahedron, accompanied with semiconductor-to-metal transition. The metallization is ascribed to the bandwidth broaden and orbital population variation. Electronic structure and electron-phonon coupling calculations imply the superconductivity in the high-pressure pyrite phase originates from the strong electron-phonon coupling assisted with topological nodal-line states. Present theoretical results not only reveal the unprecedented geometrical, mechanical and electronic properties of the bulk $PdS_2$, but also provide new insights into the interplay between a variety of structural and electronic properties, hopefully motivates further experimental and theoretical investigations of interesting physics in TMDs.

**ACKNOWLEDGMENTS**

We acknowledge Gunter Heymann and Fei Du for useful discussions. The work was sponsored by the National Natural Science Foundation of China (No. 11864008), Guangxi Natural Science Foundation (No. 2018GXNSFAA138185, 2018AD19200 and 2019GXNSFGA245006) and the Scientific Research Foundation of Guilin University of Technology (No. GUTQDJJ2017105). C.A. is supported by the Foundation for Polish Science through the International Research Agendas program co-financed by the European Union within the Smart Growth Operational Programme. High performance computational resources provided by LvLiang Cloud Computing Center of China and National Supercomputer Center on TianHe-2 are also gratefully acknowledged.

# Supplemental Information

**Structural transition, metallization and superconductivity in quasi 2D layered PdS$_2$ under compression**

Wen Lei,[1] Wei Wang,[1] Xing Ming,[1,*] Shengli Zhang,[2,†] Gang Tang,[3] Xiaojun Zheng,[1] Huan Li,[1] and Carmine Autieri[4]

   1. College of Science, Guilin University of Technology, Guilin 541004, China.

   2. MIIT Key Laboratory of Advanced Display Materials and Devices, School of Materials Science and Engineering, Nanjing University of Science and Technology, Nanjing 210094, China.

   3. Theoretical Materials Physics, Q-MAT, CESAM, University of Liège, B-4000 Liège, Belgium.

   4. International Research Centre MagTop, Institute of Physics, Polish Academy of Sciences, Aleja Lotników 32/46, PL-02668 Warsaw, Poland.

---

[*] Email: mingxing@glut.edu.cn (Xing Ming)
[†] Email: zhangslvip@njust.edu.cn (Shengli Zhang)

**Table S1** Theoretical calculated together with experimental measured equilibrium lattice constants, cell volume and positional parameters (Pd (0, 0, 0) and S (x, y, z)) for the orthorhombic PdS$_2$. The data calculated within PBE, DFT-D (TS) and DFT-D (Grimme) are from present work.

| PdS$_2$ | a (Å) | b (Å) | c (Å) | V (Å$^3$) | x | y | z |
|---|---|---|---|---|---|---|---|
| Exp.[1] | 5.460 | 5.541 | 7.531 | 227.84 | 0.107 | 0.112 | 0.425 |
| Cal.[2] | 5.465 | 5.538 | 7.525 | 227.75 | 0.104 | 0.109 | 0.416 |
| PBE | 5.494 | 5.590 | 8.633 | 265.15 | 0.104 | 0.110 | 0.427 |
| DFT-D (TS) | 5.531 | 5.600 | 7.591 | 235.12 | 0.106 | 0.110 | 0.420 |
| DFT-D (Grimme) | 5.498 | 5.574 | 7.514 | 230.25 | 0.104 | 0.109 | 0.417 |

**Table S2** Calculated bond lengths, interatomic distances and interlayer distances for the orthorhombic PdS$_2$ compared with experimental data, here d1 is the interlayer distance between Pd and S atoms, d2 and d3 are the Pd-S bond lengths in the square-planar (PdS$_4$)$^{2-}$ structural units, d4 is the S-S bond lengths in the (S$_2$)$^{2-}$ anions. The data calculated within PBE, DFT-D (TS) and DFT-D (Grimme) are from present work.

| PdS$_2$ | Exp.[1] | Cal.[2] | PBE | DFT-D (TS) | DFT-D (Grimme) |
|---|---|---|---|---|---|
| d1 (Å) | 3.312 | 3.238 | 3.782 | 3.302 | 3.245 |
| d2 (Å) | 2.300 | 2.320 | 2.340 | 2.339 | 2.336 |
| d3 (Å) | 2.304 | 2.334 | 2.348 | 2.346 | 2.345 |
| d4 (Å) | 2.130 | 2.086 | 2.096 | 2.087 | 2.081 |

**Table S3** Ferroelastic strains and activation energy barrier of different materials (ML stands for 2D monolayer).

|  | Strain (%) | activation energy barrier (meV/atom) | Ref. |
|---|---|---|---|
| Phosphorene(ML) | 37.9 | 200 | [3] |
| SnS(ML) | 4.9 | 4.2 | [3] |
| SnSe(ML) | 2.1 | 1.3 | [3] |
| GeS(ML) | 17.8 | 22.6 | [3] |
| GeSe(ML) | 6.6 | 9.8 | [3] |
| t-YN(ML) | 14.4 | 33 | [4] |
| $BP_5$(ML) | 41.4 | 320 | [5] |
| Borophane(ML) | 42 | 100 | [6] |
| TiN(ML) | 4.4 | 0.8 | [7] |
| InOCl(ML) | 17.65 | 106 | [8] |
| InOBr(ML) | 13.85 | 79 | [8] |
| MoSSe | 0.9 | 0.8 | [9] |
| WSSe | 4.7 | 3.4 | [9] |
| $PdSe_2$(bulk) | 32.3 | 22.5 | [10] |
| $PdS_2$(bulk) | 34.8 | 7.64 | present work |

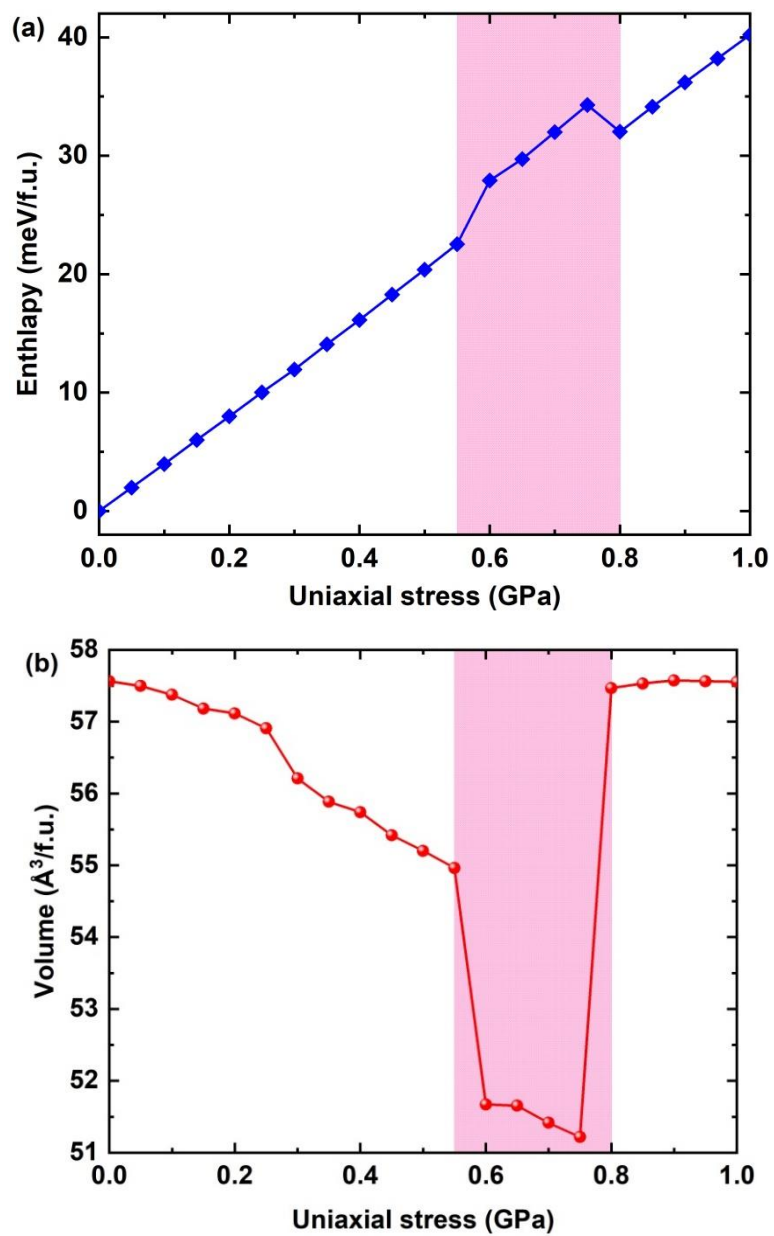

**Fig. S1** Enthalpy (a) and cell volume (b) in response to the uniaxial compressive stress along the *c*-axis. The shadow region indicates the intermediate phase.

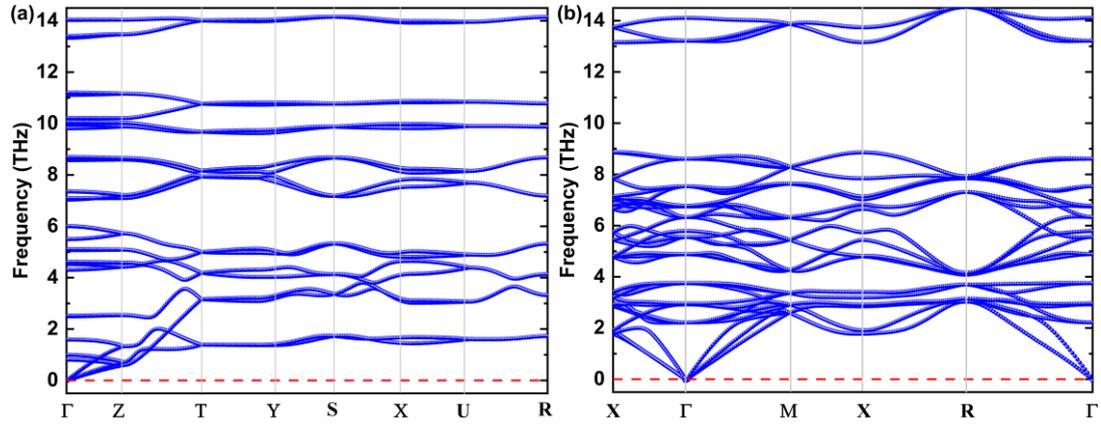

**Fig. S2** The phonon spectra calculated with a $2 \times 2 \times 2$ supercell of the PdS$_2$ for (a) *Pbca* phase and (b) *Pa$\bar{3}$* phase at 0 GPa based on the PBE level, respectively.

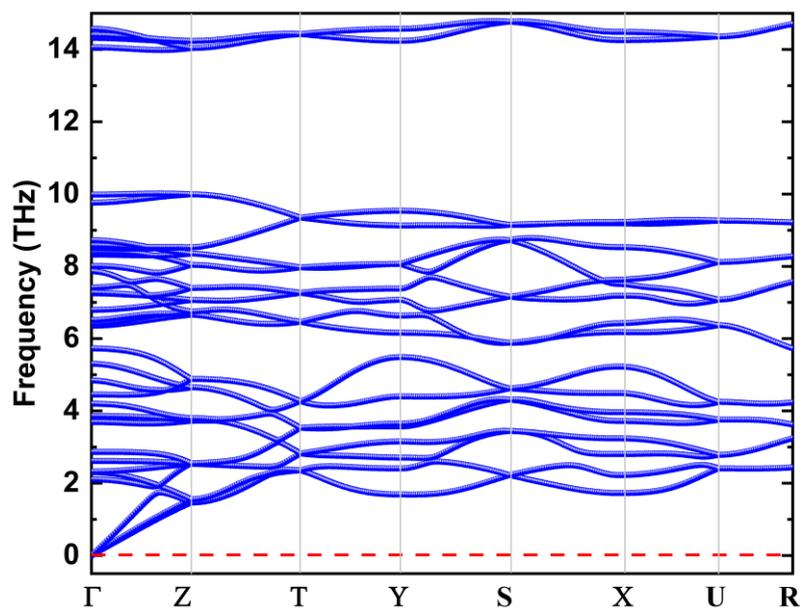

**Fig. S3** The phonon spectra calculated with a $2 \times 2 \times 2$ supercell of the intermediate phase PdS$_2$ at hydrostatic pressure of 2 GPa.

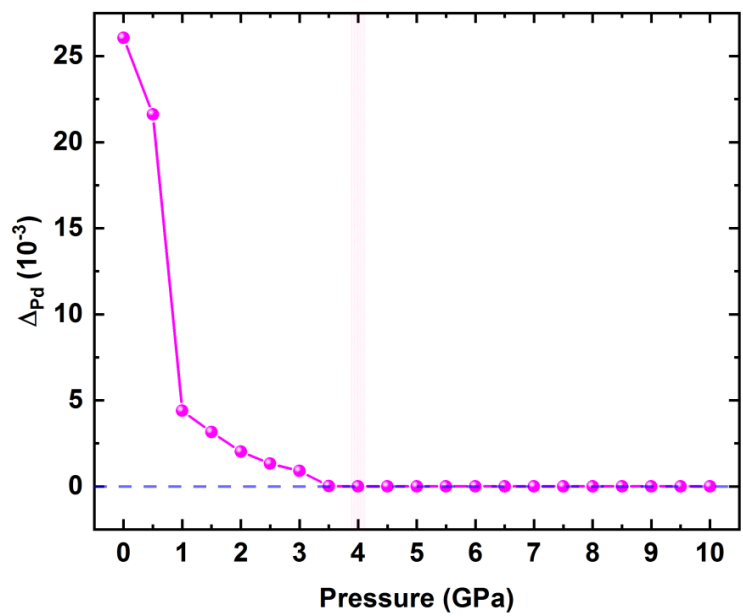

**Fig. S4** The evolution of distortive degree ($\Delta_{Pd}$) with hydrostatic pressure.

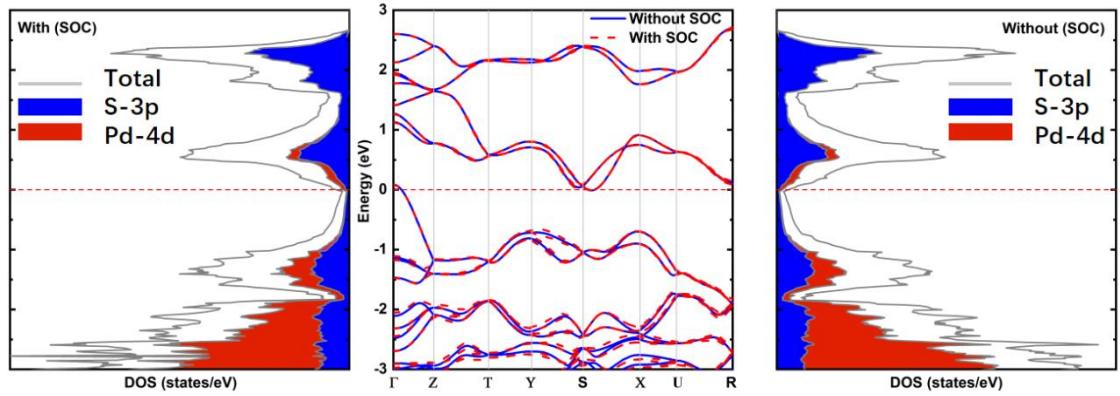

**Fig. S5** Calculated band structure and DOS of the orthorhombic PdS$_2$ with DFT-D (Grimme) without/with SOC. The Fermi energy is set to zero.

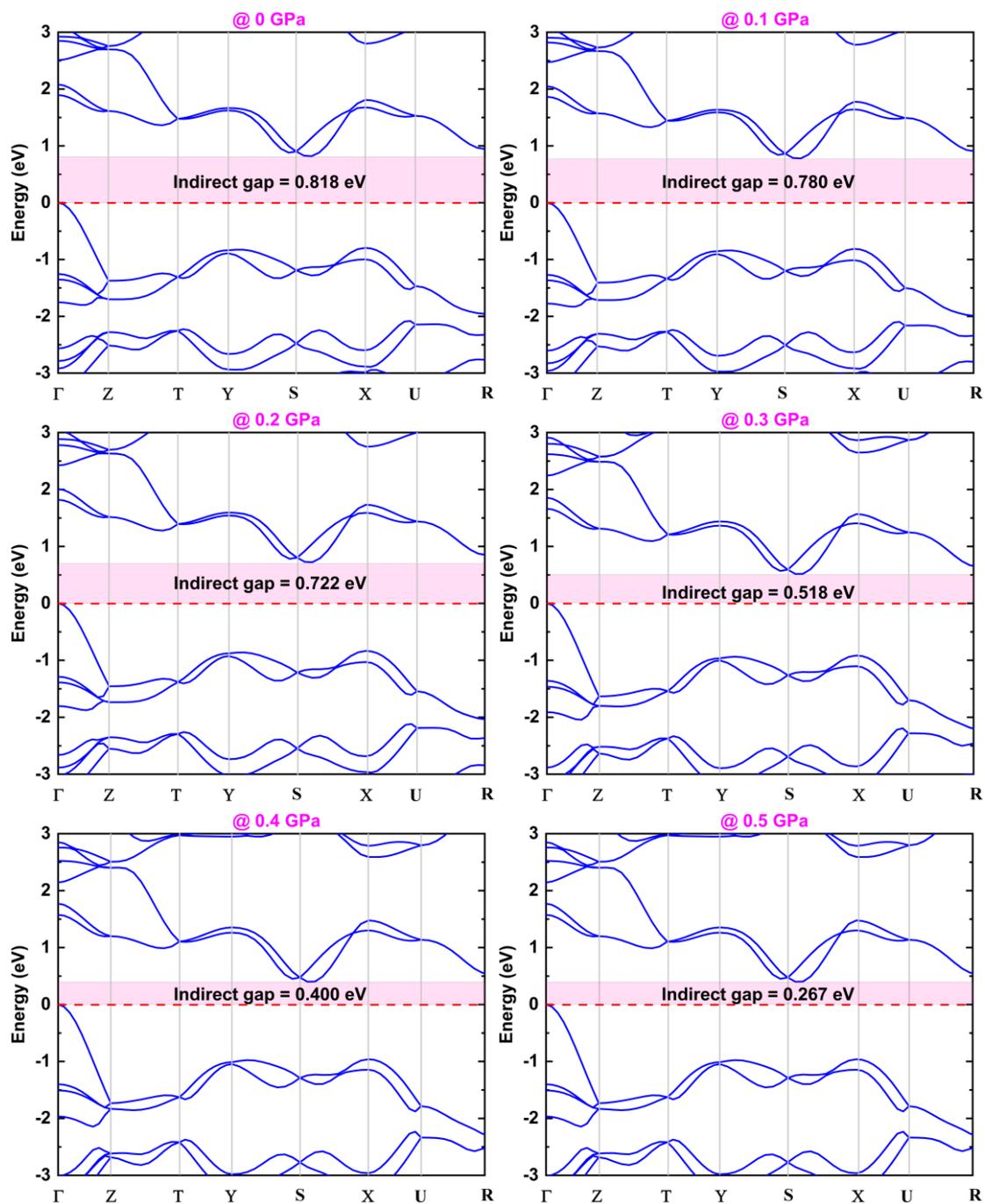

**Fig. S6** Band structures of orthorhombic PdS$_2$ under uniaxial compressive stress calculated within HSE06. The Fermi level and VBM are set to zero. In fact, real Fermi level and VBM gradually increase along with the increase of the compression stress.

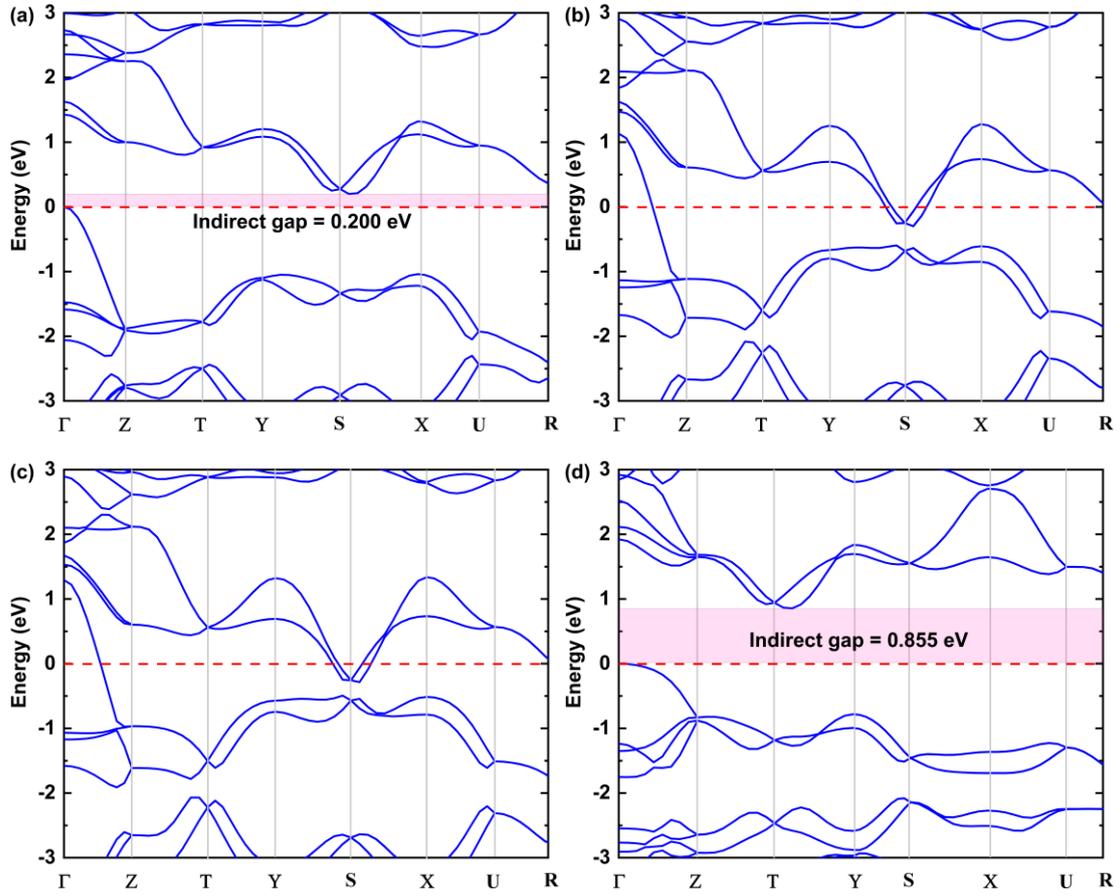

**Fig. S7** Electronic band structures of orthorhombic PdS$_2$ uniaxial compressive stress calculated within HSE06 around the ferroelastic phase transition: (a) 0.55 GPa, (b) 0.6 GPa, (c) 0.7 GPa and (d) 0.8 GPa. The Fermi level is set to zero.

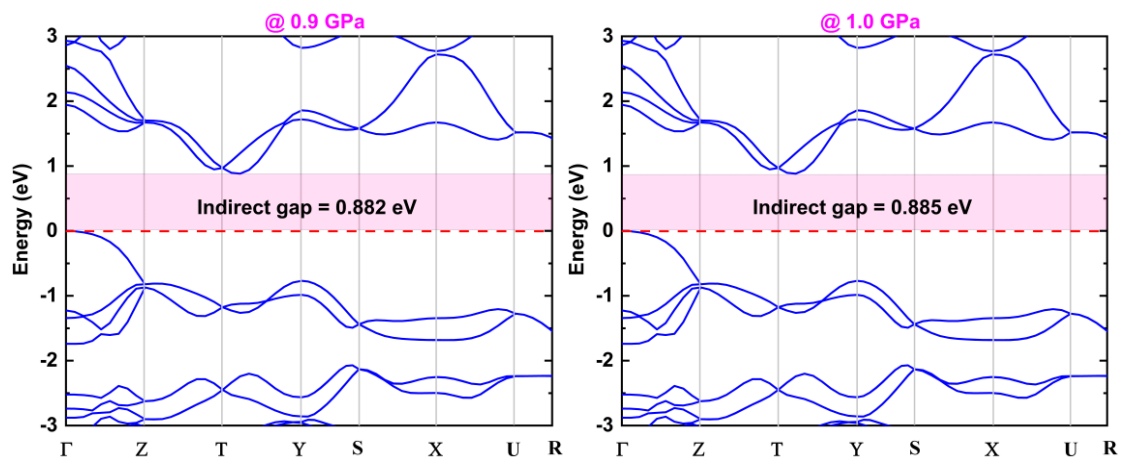

**Fig. S8** Band structure of PdS$_2$ uniaxial compressive stress calculated within HSE06 after the ferroelastic lattice rotation. The Fermi level and VBM are set to zero.

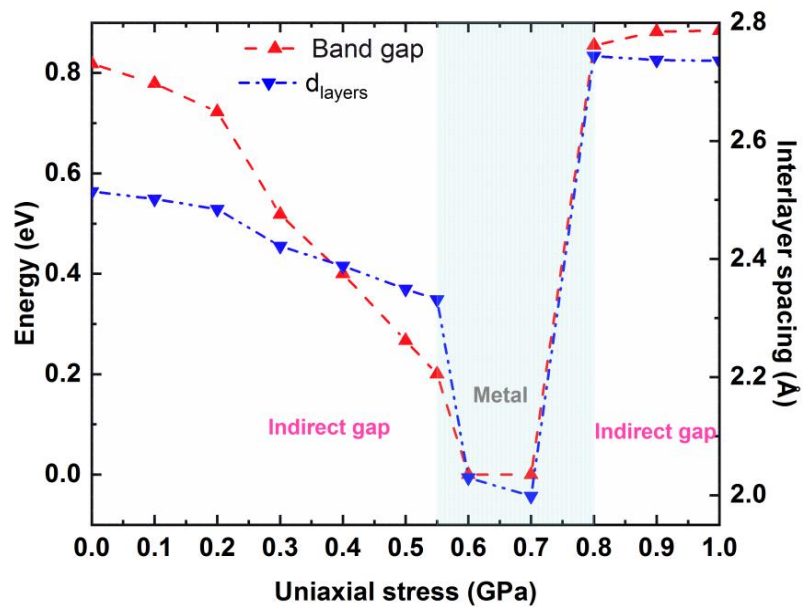

**Fig. S9** The evolution of band gap and interlayer spacing ($d_{layers}$) along with the uniaxial compressive stress.

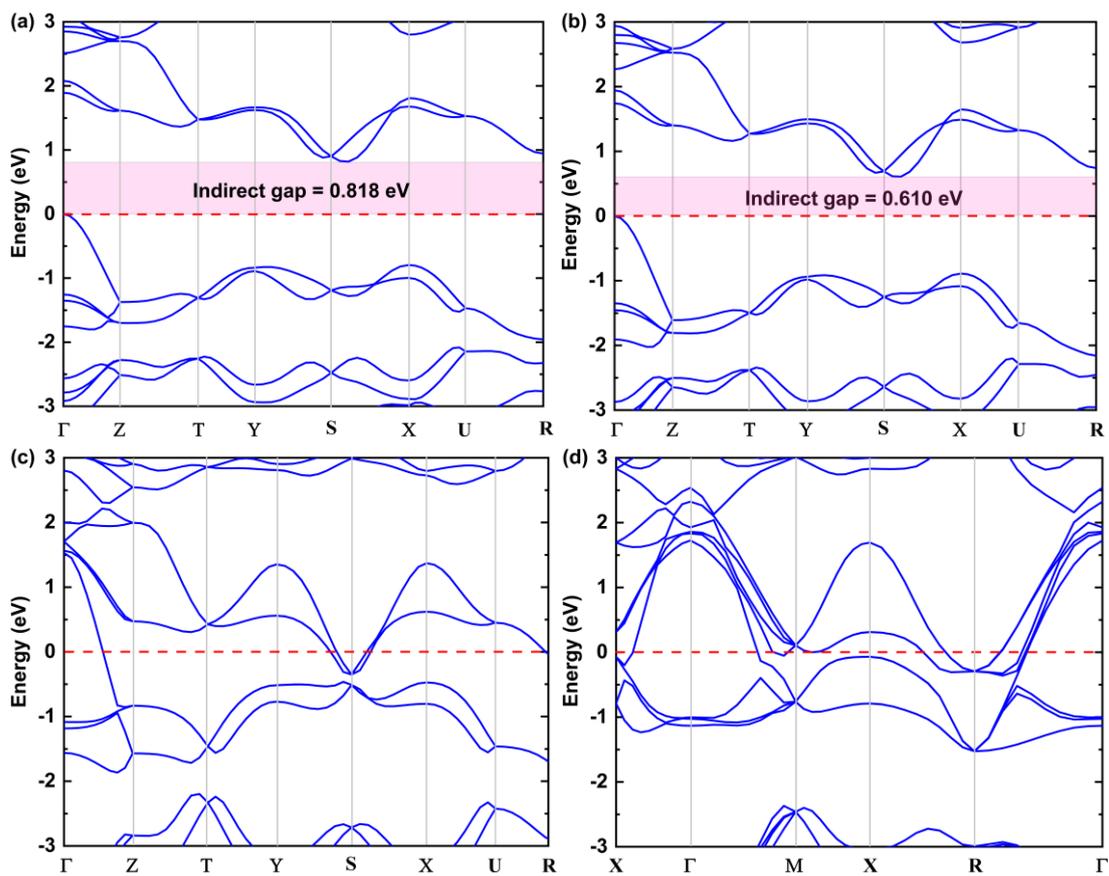

**Fig. S10** Electronic band structures of PdS$_2$ under hydrostatic pressure calculated within HSE06: (a) initial phase at 0 GPa and (b) 0.5 GPa, (c) intermediate phase at 2 GPa, and (d) final phase at 4 GPa.

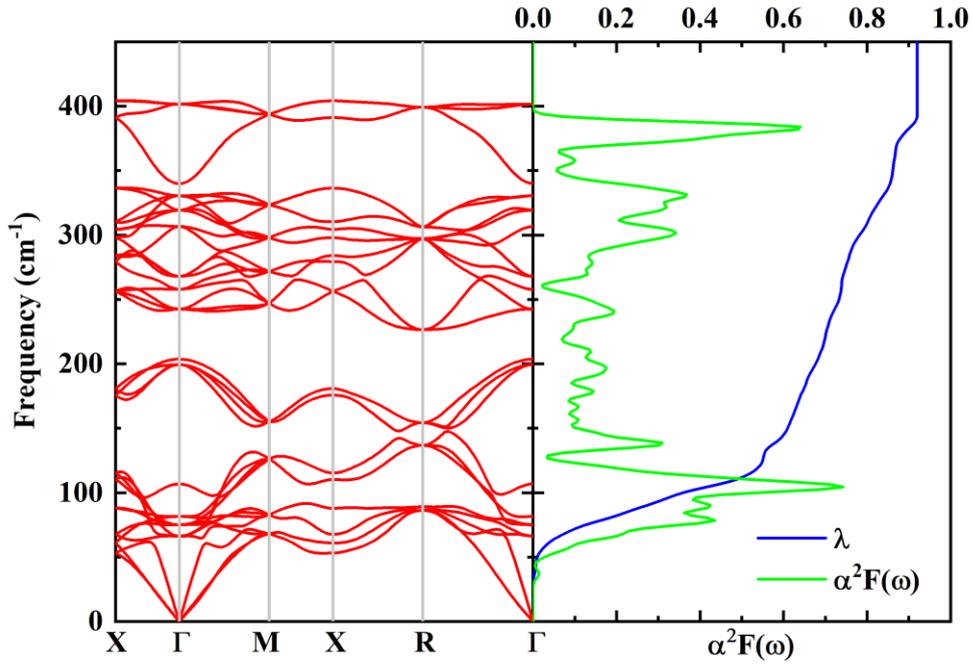

**Fig. S11** Left: phonon dispersion curves for pyrite-type PdSe$_2$ at 10 GPa. Right: Eliashberg electron-phonon coupling spectral function $\alpha^2F(\omega)$ and the electron-phonon coupling integral $\lambda(\omega)$.